\documentclass[journal]{IEEEtran}

\usepackage{epsfig,amsmath,amssymb,epsf,amsthm,scalefnt,multirow,subfig}
\usepackage{xcolor}
\usepackage{float}
\usepackage{cite}
\usepackage{psfrag}
\usepackage{booktabs}
\usepackage{algorithm}
\usepackage{algpseudocode}

\newcommand{\argmax}{\mathop{\mathrm{argmax}}}
\newcommand{\argmin}{\mathop{\mathrm{argmin}}}

\newtheorem{lemma}{Lemma}
\newtheorem*{lemma*}{Lemma}

\newtheorem{corollary}{Corollary}

\newtheorem{proposition}{Proposition}

  \def\cC{{\mathcal{C}}}

 \def\cN{{\mathcal{N}}}  
  \def\cS{{\mathcal{S}}}

\def\argmin{\mathop{\mathrm{argmin}}}
\def\argmax{\mathop{\mathrm{argmax}}}

\def\b0{{\pmb{0}}} 

\def\ba{{\mathbf{a}}} \def\bb{{\mathbf{b}}}  
\def\bee{{\mathbf{e}}}  \def\bg{{\mathbf{g}}} \def\bh{{\mathbf{h}}}
   
 \def\bn{{\mathbf{n}}}  
   
  \def\bw{{\mathbf{w}}} \def\bx{{\mathbf{x}}}
\def\by{{\mathbf{y}}} \def\bz{{\mathbf{z}}}  

\def\bA{{\mathbf{A}}}   
  \def\bG{{\mathbf{G}}} \def\bH{{\mathbf{H}}}
\def\bI{{\mathbf{I}}}

   \def\bX{{\mathbf{X}}}

\begin{document}

\title{Near Maximum-Likelihood Detector and Channel Estimator for Uplink Multiuser Massive MIMO Systems with One-Bit ADCs}

\author{Junil Choi, Jianhua Mo, and Robert W. Heath Jr.\\
\thanks{The authors are with Wireless Networking and Communications Group, The University of Texas at Austin, Austin, TX 78712, USA (email: \{junil.choi,jhmo,rheath\}@utexas.edu).}
\thanks{This work was sponsored in part by the U.S. Department of Transportation through the Data-Supported Transportation Operations and Planning (D-STOP) Tier 1 University Transportation Center and in part by the National Science Foundation under Grant No. NSF-CCF-1319556.}}

\maketitle

\begin{abstract}
In massive multiple-input multiple-output (MIMO) systems, it may not be power efficient to have a high-resolution analog-to-digital converter (ADC) for each antenna element. In this paper, a near maximum likelihood (nML) detector for uplink multiuser massive MIMO systems is proposed where each antenna is connected to a pair of one-bit ADCs, i.e., one for each real and imaginary component of the baseband signal.  The exhaustive search over all the possible transmitted vectors required in the original maximum likelihood (ML) detection problem is relaxed to formulate an ML estimation problem. Then, the ML estimation problem is converted into a convex optimization problem which can be efficiently solved.  Using the solution, the base station can perform simple symbol-by-symbol detection for the transmitted signals from multiple users. To further improve detection performance, we also develop a two-stage nML detector that exploits the structures of both the original ML and the proposed (one-stage) nML detectors. Numerical results show that the proposed nML detectors are efficient enough to simultaneously support multiple uplink users adopting higher-order constellations, e.g., 16 quadrature amplitude modulation. Since our detectors exploit the channel state information as part of the detection, an ML channel estimation technique with one-bit ADCs that shares the same structure with our proposed nML detector is also developed.  The proposed detectors and channel estimator provide a complete low power solution for the uplink of a massive MIMO system.
\end{abstract}

\section{Introduction}\label{sec1}
Massive multiple-input multiple-output (MIMO) is a transmission technique for cellular systems that leverages a large number of base station antennas to support many single antenna users \cite{Marzetta:2010}.  With enough antennas, massive MIMO can eliminate inter-user interference completely using matched beamforming for downlink and matched combining for uplink if the base station has full channel state information (CSI).  Because of its simple beamforming and combining structures, massive MIMO will be beneficial not only for cellular systems but also for other wireless communication systems, e.g., vehicular-to-everything (V2X) communications using many antennas \cite{Zhan:2013}.


There are several \textit{practical} constraints that are encountered when implementing massive MIMO systems.  Because of the large number of antennas, it may not be possible to deploy expensive and powerful hardware with small noise and distortion at the base station.  Prior work studied the impact of several hardware impairments including phase-drifts due to non-ideal oscillators and distortion noise caused by analog-to-digital convertors (ADCs) for massive MIMO systems.  It was shown in \cite{Emil:2013,Zhang:2015,Emil:2015} that having a large number of antennas helps to mitigate these hardware impairments, which confirms the benefit of massive MIMO.

In this paper, we focus on uplink multiuser massive MIMO systems using extremely low-resolution of one-bit ADCs at the base station.  Because the power consumption by ADCs grows exponentially with their resolution level \cite{Walden:1999,Murmann:2015}, using one-bit ADCs may be a practical way of implementing cost-efficient and green massive MIMO systems.  It is expected that using one-bit ADCs is particulary beneficial for wideband communication systems (when properly handling frequency selectivity) that require high sampling frequency, which will become common in millimeter wave (mmWave) communication systems \cite{Rappaport:2014}.  Adopting one-bit ADCs for uplink massive MIMO, however, is challenging because of the severe threshold applied to the received signal.

Using low-resolution ADCs for wireless communications has been investigated under various assumptions.  It was shown in \cite{Mezghani:2008} for one-bit ADCs and in \cite{Singh:2009} for low-resolution (one to three bits) ADCs that the capacity maximizing transmit signals for single-input single-output (SISO) channel are discrete, which is different from the unquantized output case.  It was shown in \cite{Wang:2013tcom} that even a low-complexity suboptimal QPSK detector with one-bit ADC suffers only 1-3 dB signal-to-noise-ratio (SNR) loss compared to the unquantized case for SISO channels. In \cite{Koch:2010,Zhang:2012tcom}, it was shown that oversampling recovers some of the loss in the SISO channel capacity incurred by one-bit quantization. The mutual information of the MIMO channel with quantized output was studied without optimizing input distribution in \cite{Ivrlac:2006} for one-bit ADCs and in \cite{Murray:2006,Nossek:2006} for low-resolution ADCs. The input distributions were optimized to maximize the achievable rate of the quantized MIMO channel using one-bit ADCs in \cite{Mezghani:2007,Mo:2015}.  Interestingly, the impact of having low-resolution ADCs is not that severe. For example, \cite{Nossek:2006} showed that the rate loss due to using one-bit ADCs is 1.5793 bits/s for 4$\times$4 MIMO with a quadrature phase shift keying (QPSK) constellation. It was proven in \cite{Mezghani:2007} that the mutual information of MIMO with one-bit ADCs is only $2/\pi$ times smaller compared to that of the unquantized MIMO case in the low SNR regime. With an optimized threshold, this gap can vanish in the low SNR regime \cite{Koch:2013}. In \cite{Orhan:2015}, digital and analog combiners using low-resolution ADCs were compared in terms of the achievable rate for point-to-point mmWave systems where the derived achievable rates were tight only for the low SNR regime.

The aforementioned work on quantized MIMO \cite{Ivrlac:2006,Murray:2006,Nossek:2006,Mezghani:2007,Mo:2015,Koch:2013,Orhan:2015} was restricted to point-to-point communications and focused on studying capacity or designing optimal transmit signals for quantized channel outputs using low-resolution ADCs without proposing specific signal processing algorithms for detecting the received signals. For point-to-point communications with low, but more than one-bit resolution (e.g., 2-3 bits) ADCs, a modified minimum mean square error (MMSE) detector was proposed in \cite{Mezghani:2007b} and later extended to an iterative decision feedback equalizer in \cite{Mezghani:2012b}. It is possible to extend the iterative detection technique developed in \cite{Mezghani:2012b} to multiuser scenarios. The iterative detector proposed in \cite{Mezghani:2012b} exploits the assumption of nearest neighbor error of adopted constellation, which may not be effective for practical channel models.

In this paper, we propose a practical near maximum likelihood (nML) detector for uplink multiuser massive MIMO systems with one-bit ADCs.  The complexity of the maximum likelihood (ML) detector grows exponentially with the number of users, which makes it difficult to use for massive MIMO with multiple users.  The proposed nML detector, however, can be implemented with standard convex optimization techniques with marginal performance degradation when the number of receive antennas at the base station is large, and supports multiple users with arbitrary constellation sizes. To further improve performance, we also propose a two-stage nML detector, which exploits the structures of both the original ML detector and the one-stage nML detector. Numerical results show that, with marginal additional complexity, the two-stage nML detector can significantly improve the detection performance and achieve almost the same performance with the original ML detector.

There are several recent papers on massive MIMO with low-resolution ADCs. It was shown in  \cite{Risi:2014} and \cite{Jacobsson:2015} that linear detectors with one-bit ADCs work well for multiuser scenarios with QPSK and 16 quadrature amplitude modulation (QAM) constellations, respectively. An uplink multiuser massive MIMO detector was developed in \cite{Wang:2015TWC} but the detector was based on several bit ADCs and developed for the spatial modulation transmission technique \cite{Renzo:2014}. For general symbol transmission in uplink massive MIMO systems with one-bit ADCs, a message-passing algorithm-based multiuser detector was proposed in \cite{Wang:2014} for special constellation structures. The detector was later extended to arbitrary constellations in \cite{Wang:2015icc}. Our numerical studies show that our nML detector works well with both perfect CSI and channel direction information (CDI) while the detector in \cite{Wang:2015icc} experiences the performance degradation with CDI when SNR is low. In \cite{Wang:2014icc}, a multiuser detector using low-resolution ADCs was developed based on convex optimization, which has a similar structure with our nML detector. The convex optimization process in \cite{Wang:2014icc}, however, is optimized separately for each constellation while our nML detector is able to detect arbitrary constellations without any modification. The mixed use of one-bit and high-resolution ADCs in massive MIMO was analyzed in \cite{Liang:2015}, where it was found that using one-bit ADCs with a few high-resolution ADCs can achieve similar performance with the unquantized case. Joint channel and data estimation using low-resolution ADCs was proposed in \cite{Wen:2015}. While the performance was comparable to the case with perfect CSI, the computational complexity of the joint technique \cite{Wen:2015} may still be too high to be affordable in a commercial system.

Since our detector exploits CSI as part of the detection, we also propose an ML channel estimation technique using one-bit ADCs that is in line with our proposed nML detectors.  There has been related work on channel estimation with low-resolution ADCs \cite{Dabeer:2010} and with one-bit ADCs \cite{Lok:1998,Ivrlac:2007,Mezghani:2012,Mo:2014Asilomar,Choi:2014Asilomar}.  In \cite{Lok:1998,Ivrlac:2007}, the expectation-maximization (EM) algorithm was exploited to estimate channels. The problem solved in \cite{Lok:1998,Ivrlac:2007}, however, is not convex, and the EM algorithm may converge to a local optimal in the high SNR regime. The work \cite{Mezghani:2012,Mo:2014Asilomar} used generalized approximate message passing (GAMP) algorithms for channel estimation with one-bit ADCs where the techniques heavily relied on the sparsity of the target vector. While the work in \cite{Ivrlac:2007,Mo:2014Asilomar} was not able to estimate the norm of the channel due to the simple zero-threshold setting for one-bit quantization, \cite{Mezghani:2012} adopted asymmetric one-bit quantizers to deliver the the norm information of the target vector. Our ML channel estimator, which is an extension of \cite{Choi:2014Asilomar}, is shown to estimate not only the channel direction but also the channel norm and does not make an assumption about sparsity in the channel.

Our contributions are summarized as follows.
\begin{itemize}
    \item
    We propose an nML detector for uplink multiuser massive MIMO systems.  The proposed nML detector is based on the ML detector developed for distributed reception in \cite{Choi:2015}.  We show that the two problems, i.e., distributed reception and multiuser detection, are essentially the same problem, which makes it possible to exploit the detectors from \cite{Choi:2015} for our multiuser detection problem.  The complexity of the ML detector in \cite{Choi:2015}, however, grows exponentially with the number of uplink users, which prevents its use in massive MIMO with many users.  We reformulate the ML detector in \cite{Choi:2015} to derive an ML estimator and convert the ML estimation problem into a convex problem.  Therefore, we can rely on efficient convex optimization techniques to obtain an ML estimate and perform symbol-by-symbol detection based on the ML estimate.

	\item
	We implement a two-stage nML detector to further improve the detection performance. The two-stage nML detector exploits the structure of the original ML detector with the reduced candidate set constructed by the proposed (one-stage) nML detector. Numerical results show that the two-stage nML detector improves the detection performance significantly in the high SNR regime and achieves similar performance with the original ML detector.
	
    \item
	 We derive the exact probability that two different transmit vectors (after they have passed through the channel and noise is added) result in the same quantized received signal with one-bit ADCs in a certain condition. The result shows the relationship between the probability and the numbers of antennas and users, where the probability of having the same quantized outputs goes to zero as the number of antennas goes to infinity.

    \item
    We propose an ML channel estimator that has the same structure with the proposed nML detector.  The proposed estimator can estimate the direction \textit{and} norm of the channel more accurately than other channel estimators using one-bit ADCs.  Because of the similar structure, it is possible to implement both the nML detector and the ML channel estimator using the same algorithm.
\end{itemize}

The paper is organized as follows.  We describe our system model using one-bit ADCs in Section \ref{sec2}.  In Section \ref{sec3}, we briefly discuss the detectors which were originally developed for distributed reception in \cite{Choi:2015}.  Then, we propose our nML detectors and present the asymptotic analyses in Section \ref{sec4}.  We propose an ML channel estimator in Section~\ref{sec_channel_est}.  In Section~\ref{simul}, we evaluate the proposed techniques by simulations, and the conclusion follows in Section \ref{conclusion}.

\textbf{Notation:} Lower and upper boldface letters represent column vectors and matrices, respectively.  $\|\ba\|$ is used to denote the $\ell_2$-norm of a vector $\ba$, and $\bA^{\mathrm{T}}$, $\bA^{*}$, $\bA^\dagger$ denote the transpose, Hermitian transpose, and pseudo inverse of the matrix $\bA$, respectively.  $\mathrm{Re}(\bb)$ and $\mathrm{Im}(\bb)$ represent the real and complex part of a complex vector $\bb$, respectively.  $\mathbf{0}_m$ is used for the $m\times 1$ all zero vector, and $\bI_m$ denotes the $m \times m$ identity matrix.  $\mathbb{C}^{m\times n}$ and $\mathbb{R}^{m\times n}$ represent the set of all $m \times n$ complex and real matrices, respectively.

\section{System Model}\label{sec2}
We explain our system model and several assumptions that are relevant to our detector design.  We also define expressions for one-bit ADCs in this section.

\subsection{Massive MIMO Received Signal Model }

We consider a uplink multiuser cellular system with  $N_{\mathrm{c}}$ cells.  Each cell consists of a base station with $N_{\mathrm{r}}$ received antennas and $K$ users equipped with a single transmit antenna.  All $KN_{\mathrm{c}}$ users transmit independent data symbols simultaneously to their serving base stations.  Assuming all users transmit data with power $P$, the received signal at the $i$-th base station $\by_i=\begin{bmatrix}y_{1,i} & y_{2,i} & \cdots & y_{N_{\mathrm{r}},i}\end{bmatrix}^{\mathrm{T}}$ is
\begin{equation}\label{input_output1}
  \by_i = \sqrt{P}\sum_{k=1}^{K}\bh_{i,ik} x_{ik}+\sqrt{P}\sum_{\substack{m=1 \\ m\neq i}}^{N_{\mathrm{c}}}\sum_{k=1}^{K}\bh_{i,mk} x_{mk}+\bn_{i}
\end{equation}
where $\bh_{i,mk}\in \mathbb{C}^{N_{\mathrm{r}}\times 1}$ is the channel vector between the $i$-th base station and the $k$-th user associated with the $m$-th base station, $x_{mk}$ is the data symbol, which satisfies $\mathbb{E}\left[x_{mk}\right]=0$ and $\mathbb{E}\left[|x_{mk}|^2\right]=1$, transmitted from the $k$-th user supported by the $m$-th base station, and $\bn_i\sim \cC\cN(\mathbf{0}_{N_{\mathrm{r}}},\sigma^2\bI_{N_{\mathrm{r}}})$ is the complex additive white Gaussian noise (AWGN) at the $i$-th base station.

Before developing the detectors, we make the following assumptions.

\textbf{Assumption 1:} The data symbols $x_{ik}$ are from an $M$-ary constellation $\cS=\{s_1,\cdots,s_M\}$ and have been normalized such that
\begin{equation}\label{norm_const}
  \|\bx_i\|^2=K
\end{equation}
for all $i$ where $\bx_i=\begin{bmatrix}x_{i1} & x_{i2} & \cdots & x_{iK}\end{bmatrix}^{\mathrm{T}}$.  If $\cS$ is a phase shift keying (PSK) constellation satisfying $|s_m|^2=1$ for all $m$, then the norm constraint is trivially satisfied.  For a QAM constellation, the law of large numbers (LLN) gives
\begin{equation}\label{norm_approx}
  \sum_{k=1}^{K} |x_{ik}|^2 \approx K
\end{equation}
as $K$ becomes large if all users select their data symbols independently and with equal probability within $\cS$ (with proper normalization).  Although the norm constraint in \eqref{norm_const} plays an important role in implementing the proposed nML detector, numerical results in Section \ref{simul} shows that the nML detector works well even with moderate numbers of users.

\textbf{Assumption 2:} We assume that each base station has perfect local CSI with which to implement its detector.  After implementing the detectors, we relax this assumption in Section~\ref{sec_channel_est} where we consider channel estimation techniques for one-bit ADCs. We neglect pilot contamination during the channel estimation procedure because the channel is already severely distorted due to one-bit ADCs. Note that pilot contamination, which contaminates the channel estimate, remains the only channel impairment with full-resolution ADCs and extremely large number of antennas \cite{Gopalakrishnan:2011,Hoydis:2013}.

In addition to these assumptions, we first focus on the single cell scenario because the detectors considered in this paper do not exploit any kind of inter-cell cooperation.  In Section \ref{multicell_sec}, we explain how the proposed detector can be adapted to a multicell setting.

For the single cell scenario, the received signal $\by$ in \eqref{input_output1} becomes
\begin{align}\label{input_output2}
  \by = \sqrt{P}\sum_{k=1}^{K}\bh_{k} x_{k}+\bn=\sqrt{P}\bH\bx+\bn,
\end{align}
and the SNR is
\begin{equation}
  \rho=\frac{P}{\sigma^2}.
\end{equation}
We assume the base station knows $\rho$ perfectly.

\subsection{Received Signal Representation with One-Bit ADCs}
\begin{figure}[t]
\centering
  \psfrag{a}[][][0.8]{$K$ users}
  \psfrag{b}[][][0.8]{$\bH$}
  \psfrag{c}[][][0.8]{$N_{\mathrm{r}}$}
  \psfrag{d}[][][0.5]{$\mathrm{Re}(y_1)$}
  \psfrag{e}[][][0.5]{$\mathrm{Im}(y_1)$}
  \psfrag{f}[][][0.5]{$\mathrm{Re}(y_{N_{r}})$}
  \psfrag{g}[][][0.5]{$\mathrm{Im}(y_{N_{r}})$}
  \psfrag{h}[][][0.8]{$\hat{y}_1$}
  \psfrag{i}[][][0.8]{$\hat{y}_{N_{\mathrm{r}}}$}
\includegraphics[width=1\columnwidth]{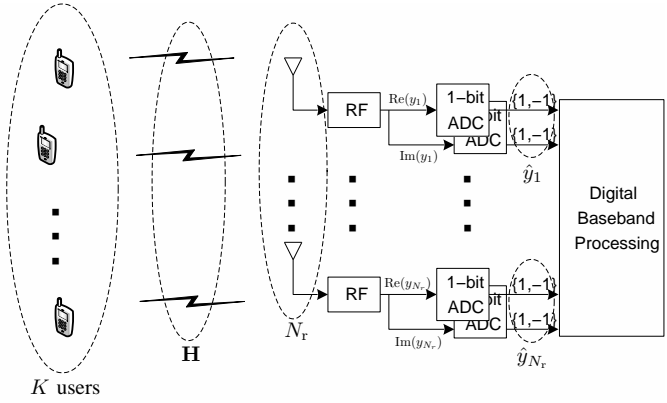}
\caption{MU-MIMO with $K$ users and $N_{\mathrm{r}}$ receive antennas.  Each received signal $y_n$ is processed with two one-bit ADCs.}\label{concept_fig}
\end{figure}

We focus on a massive MIMO system that uses one-bit ADCs for the real and imaginary parts of the each element of $\by$.  The conceptual figure of our system is depicted in Fig. \ref{concept_fig}.

The output of the $n$-th receive antenna after the one-bit ADCs is given as
\begin{equation}
  \hat{y}_n=\mathrm{sgn}(\mathrm{Re}(y_n))+j\mathrm{sgn}(\mathrm{Im}(y_n))
\end{equation}
where $\mathrm{sgn}(\cdot)$ is the sign function which is defined as
\begin{equation}
\mathrm{sgn}(x) = \begin{cases}1& \text{if } x\geq 0 \\
-1 & \text{if } x<0\end{cases}.
\end{equation}
Therefore, we have
\begin{equation}
  \hat{y}_n\in \left\{1+j, -1+j, -1-j, 1-j\right\}
\end{equation}
for $1\leq n \leq N_{\mathrm{r}}$. The collection of $\hat{y}_n$ is given as
\begin{align}
  \hat{\by}=\begin{bmatrix}\hat{y}_1 & \hat{y}_2 & \cdots & \hat{y}_{N_{\mathrm{r}}}\end{bmatrix}^{\mathrm{T}}.
\end{align}

\section{Possible Detectors Using One-Bit ADCs}\label{sec3}

In this section, we reformulate the detectors for distributed reception proposed in \cite{Choi:2015} into the uplink multiuser massive MIMO setting.  We also discuss their characteristics and limitations.  This discussion is useful for developing our detectors in Section \ref{sec4}.

\subsection{ML Detector Reformulation}\label{ML_orig}

Let $\bg_n^{\mathrm{T}}\in \mathbb{C}^{1\times K}$ be the $n$-th row of the channel matrix $\bH$, i.e.,
\begin{equation}
  \bH=\begin{bmatrix}\bg_1 & \bg_2 & \cdots & \bg_{N_{\mathrm{r}}}\end{bmatrix}^{\mathrm{T}}.
\end{equation}
Note that $\bg_n$ is the channel between the $K$ users and the $n$-th receive antenna.  Assuming the real and imaginary components of the Gaussian noise are IID, it is useful when approaching one-bit ADC problems to rewrite the signal model in the real vector form instead of complex form as
\begin{align}
\bG_{\mathrm{R},n}&=\begin{bmatrix} \mathrm{Re}(\bg_n) & \mathrm{Im}(\bg_n) \\ -\mathrm{Im}(\bg_n) & \mathrm{Re}(\bg_n)\end{bmatrix}^{\mathrm{T}}=\begin{bmatrix}\bg_{\mathrm{R},n,1}^{\mathrm{T}} \\ \bg_{\mathrm{R},n,2}^{\mathrm{T}} \end{bmatrix}\in \mathbb{R}^{2\times 2K},\\
\bn_{\mathrm{R},n}&=\begin{bmatrix} \mathrm{Re}(n_n) \\ \mathrm{Im}(n_n)\end{bmatrix}=\begin{bmatrix} n_{\mathrm{R},n,1} \\ n_{\mathrm{R},n,2}\end{bmatrix}\in \mathbb{R}^{2\times 1},\\
\bx_{\mathrm{R}}&=\begin{bmatrix} \mathrm{Re}(\bx) \\ \mathrm{Im}(\bx)\end{bmatrix}\in \mathbb{R}^{2K\times 1}
\end{align}
where
\begin{equation}
  \bg_{\mathrm{R},n,1}=\begin{bmatrix}\mathrm{Re}(\bg_n) \\ -\mathrm{Im}(\bg_n)	\end{bmatrix},\quad \bg_{\mathrm{R},n,2}=\begin{bmatrix}\mathrm{Im}(\bg_n) \\ \mathrm{Re}(\bg_n)	\end{bmatrix}
\end{equation}
and
\begin{equation}\label{w_R_dist}
  n_{\mathrm{R},n,i}\sim \cN\left(0,\frac{\sigma^2}{2}\right)
\end{equation}
for all $n$ and $i$.  The received signal at the $n$-th receive antenna can be also rewritten as
\begin{align}\label{data_phase}
\by_{\mathrm{R},n}&=\begin{bmatrix} \mathrm{Re}(y_n) \\ \mathrm{Im}(y_n)\end{bmatrix}=\begin{bmatrix} y_{\mathrm{R},n,1} \\ y_{\mathrm{R},n,2}\end{bmatrix}=\sqrt{P}\bG_{\mathrm{R},n} \bx_{\mathrm{R}} + \bn_{\mathrm{R},n},
\end{align}
and the vectorized version of the quantized $\hat{y}_k$ in the real domain is given as
\begin{equation}
\hat{\by}_{\mathrm{R},n}=\begin{bmatrix} \mathrm{sgn}(\mathrm{Re}(y_n)) \\ \mathrm{sgn}(\mathrm{Im}(y_n))\end{bmatrix}=\begin{bmatrix} \hat{y}_{\mathrm{R},n,1} \\ \hat{y}_{\mathrm{R},n,2}\end{bmatrix}.\label{real_quantized}
\end{equation}
Based on $\hat{\by}_{\mathrm{R},n}$, the base station generates the \textit{sign-refined} channel matrix for the $n$-th receive antenna as
\begin{equation}
\widetilde{\bG}_{\mathrm{R},n}=\begin{bmatrix}\widetilde{\bg}_{\mathrm{R},n,1}^{\mathrm{T}} \\ \widetilde{\bg}_{\mathrm{R},n,2}^{\mathrm{T}} \end{bmatrix}
\end{equation}
where $\widetilde{\bg}_{\mathrm{R},n,i}$ is defined as
\begin{align}\label{sign_refine}
  \widetilde{\bg}_{\mathrm{R},n,i}=\hat{y}_{\mathrm{R},n,i}\bg_{\mathrm{R},n,i}.
\end{align}
Note that $\hat{y}_{\mathrm{R},n,i}=\pm 1$ depending on the sign of $y_{\mathrm{R},n,i}$.  Define $\cS_{\mathrm{R}}$ to be
\begin{equation}
  \cS_{\mathrm{R}}=\left\{\begin{bmatrix}\mathrm{Re}(s_1) \\ \mathrm{Im}(s_1)\end{bmatrix},\cdots,\begin{bmatrix}\mathrm{Re}(s_M) \\ \mathrm{Im}(s_M)\end{bmatrix}\right\}
\end{equation}
where $M$ is the size of the data symbol constellation $\mathcal{S}$.

With these definitions and using similar logic from \cite{Choi:2015}, the ML detector can be defined as
\begin{align}\label{ml_detect}
\hat{\bx}_{\mathrm{R,ML}}=\argmax_{\acute{\bx}_{\mathrm{R}}\in\cS_{\mathrm{R}}^{K}} \prod_{i=1}^{2}\prod_{n=1}^{N_{\mathrm{r}}}\Phi\left(\sqrt{2\rho}\widetilde{\bg}_{\mathrm{R},n,i}^{\mathrm{T}}\acute{\bx}_{\mathrm{R}}\right)
\end{align}
where $\Phi(t)=\int_{-\infty}^{t}\frac{1}{\sqrt{2\pi}}e^{-\frac{\tau^2}{2}}d\tau$ and $\cS_{\mathrm{R}}^{K}$ is the $K$-ary Cartesian product set of $\cS_{\mathrm{R}}$, which is ordered with the real parts of the constellations first and the imaginary parts later.  The $\sqrt{2}$ term in \eqref{ml_detect} comes from the distribution of $n_{\mathrm{R},n,i}$ given in \eqref{w_R_dist}.

\subsection{ZF-Type Detector Reformulation}
By brute-force search, the complexity of the ML detector in \eqref{ml_detect} is $M^K$ which grows exponentially with the number of users $K$.  To support a large number of $K$, the ZF-type detector was proposed in \cite{Choi:2015}.  The base station first obtains the ZF estimate as
\begin{equation}
  \check{\bx}_{\mathrm{ZF}}=\bH^{\dagger}\hat{\by}.
\end{equation}
Because the norm square of $\check{\bx}_{\mathrm{ZF}}$ may not equal to $K$, the base station normalizes $\check{\bx}_{\mathrm{ZF}}$ as
\begin{equation}
  \bar{\bx}_{\mathrm{ZF}}=\sqrt{K}\frac{\check{\bx}_{\mathrm{ZF}}}{\|\check{\bx}_{\mathrm{ZF}}\|}
\end{equation}
and performs symbol-by-symbol detection using $\bar{\bx}_{\mathrm{ZF}}$ as
\begin{equation}\label{zf_sym_detect}
  \hat{x}_{\mathrm{ZF},k}=\argmin_{\acute{x}\in \cS}|\bar{x}_{\mathrm{ZF},k}-\acute{x}|^2
\end{equation}
where $\bar{x}_{\mathrm{ZF},k}$ is the $k$-th element of normalized $\bar{\bx}_{\mathrm{ZF}}$.  The normalization is not an issue for PSK constellations; however, it is crucial for QAM constellations.

It was shown numerically in \cite{Choi:2015} that ZF-type detectors saturate at a higher error rate floor than ML detectors as SNR increases; the error rate floors in both cases are inevitable due to the one-bit ADCs. Therefore, we propose new detectors that outperform the ZF-type detector and require less complexity than the ML detector.

\section{Near ML Detector Implementation}\label{sec4}
We derive our nML detector by converting the original ML detection problem as a convex optimization problem.  This can be done by relaxing constraints on the transmitted vector.  We also analyze the nML detector in the asymptotic regimes and derive the exact probability that two different transmit vectors result in the same quantized signal (after they have passed through the channel and noise is added) in a certain condition. The relationship between the probability and the number of receive antennas show that the probability goes to zero as the number of antennas goes to infinity, which shows the benefit of massive MIMO for one-bit ADCs. Then we implement the two-stage nML detector to further improve the detection performance. Finally, we extend the proposed detectors to a multicell scenario.

\subsection{Convex Optimization Formulation of nML Detector}\label{convex_ml}
Because of the norm constraint \eqref{norm_const}, we define the ML estimator by relaxing the constraint $\acute{\bx}_{\mathrm{R}}\in\cS_{\mathrm{R}}^{K}$ in \eqref{ml_detect} as
\begin{align}
 \check{\bx}^{(1)}_{\mathrm{R,ML}}&=\argmax_{\substack{\acute{\bx}_{\mathrm{R}}\in\mathbb{R}^{2K \times 1}\\ \|\acute{\bx}_{\mathrm{R}}\|^2=K}} \prod_{i=1}^{2}\prod_{n=1}^{N_{\mathrm{r}}}\Phi\left(\sqrt{2\rho}\widetilde{\bg}_{\mathrm{R},n,i}^{\mathrm{T}}\acute{\bx}_{\mathrm{R}}\right)\\
&=\argmax_{\substack{\acute{\bx}_{\mathrm{R}}\in\mathbb{R}^{2K \times 1}\\ \|\acute{\bx}_{\mathrm{R}}\|^2=K}} \sum_{i=1}^{2}\sum_{n=1}^{N_{\mathrm{r}}}\log\Phi\left(\sqrt{2\rho}\widetilde{\bg}_{\mathrm{R},n,i}^{\mathrm{T}}\acute{\bx}_{\mathrm{R}}\right).\label{ml_est}
\end{align}
It was shown in \cite{Choi:2015} that
\begin{equation}\label{conv_lemma}
\check{\bx}^{(1)}_{\mathrm{R,ML}}\rightarrow \bx_{\mathrm{R}}
\end{equation}
in probability as $N_{\mathrm{r}}\rightarrow \infty$ for arbitrary $\rho>0$.  Therefore, if \eqref{ml_est} can be solved, then the detector is guaranteed to achieve good performance when $N_{\mathrm{r}}$ is large. In fact, all of the properly designed detectors including linear detectors become optimal as $N_{\mathrm{r}} \rightarrow \infty$ \cite{Marzetta:2010,Choi:2015}, and \eqref{conv_lemma} proves that the ML detector based on \eqref{ml_est} is indeed a proper detector.

The maximization in \eqref{ml_est} is still not easy to solve in general.  The function $\Phi(\cdot)$ is log-concave but the optimization problem in \eqref{ml_est} is not convex due to the norm constraint $\|\acute{\bx}_{\mathrm{R}}\|^2=K$.  To sidestep this challenge, we relax the norm constraint as $\|\acute{\bx}_{\mathrm{R}}\|^2\leq K$ and reformulate the problem as
\begin{align}
\check{\bx}^{(2)}_{\mathrm{R,ML}}=\argmax_{\substack{\acute{\bx}_{\mathrm{R}}\in\mathbb{R}^{2K \times 1}\\ \|\acute{\bx}_{\mathrm{R}}\|^2\leq K}} \sum_{i=1}^{2}\sum_{n=1}^{N_{\mathrm{r}}}\log\Phi\left(\sqrt{2\rho}\widetilde{\bg}_{\mathrm{R},n,i}^{\mathrm{T}}\acute{\bx}_{\mathrm{R}}\right)\label{ml_est_convex}
\end{align}
which is a convex optimization problem that can be efficiently solved \cite{Boyd:2004}. Similar to \cite{Zymnis:2010,Wang:2014icc}, we provide a simple yet effective iterative approach to solve \eqref{ml_est_convex} in Algorithm \ref{Al:01}.

\begin{algorithm}[t]
  \caption{Iterative algorithm to solve \eqref{ml_est_convex}}
  \label{Al:01}
  \begin{algorithmic}
\State \textbf{Initialization}
\State 1:~Set the initial point $\acute{\bx}_{\mathrm{R}}^{(0)}$
\State 2:~Set \\
~~~~$\widetilde{\bG}_{\mathrm{R}}=\begin{bmatrix}\widetilde{\bg}_{\mathrm{R},1,1} & \cdots & \widetilde{\bg}_{\mathrm{R},1,N_{\mathrm{r}}} & \widetilde{\bg}_{\mathrm{R},2,1} & \cdots & \widetilde{\bg}_{\mathrm{R},2,N_{\mathrm{R}}}   \end{bmatrix}^{\mathrm{T}}$
\State 3:~Set the step size $\kappa$ and the termination threshold $\epsilon$
\State \textbf{Iterative update}
\State 4:~~\textbf{While}~$\left\lVert \acute{\bx}_{\mathrm{R}}^{(k)}-\acute{\bx}_{\mathrm{R}}^{(k-1)}\right\rVert\geq \epsilon \left\lVert \acute{\bx}_{\mathrm{R}}^{(k-1)}\right\rVert$
\State 5:~~~~~$\acute{\bx}_{\mathrm{R}}^{(k)}=\acute{\bx}_{\mathrm{R}}^{(k-1)}+\kappa \widetilde{\bG}_{\mathrm{R}}^{\mathrm{T}}\triangledown f\left(\acute{\bx}_{\mathrm{R}}^{(k-1)}\right)$ \\
~~~~~~~where the $i$-th element of $\triangledown f\left(\bz\right)$ is\\
~~~~~~~$\triangledown f\left(\bz\right)_i=\frac{1}{\sqrt{2\pi}}\frac{e^{-\rho\left|\widetilde{\bg}_{\mathrm{R},n,i}^{\mathrm{T}}\bz\right|^2}}{\Phi\left(\sqrt{2\rho}\widetilde{\bg}_{\mathrm{R},n,i}^{\mathrm{T}}\bz\right)}$
\State 6:~~~~~\textbf{If}~$\left\lVert\acute{\bx}_{\mathrm{R}}^{(k)}\right\rVert^2>K$
\State 7~~~~~~~~~$\acute{\bx}_{\mathrm{R}}^{(k)}=\sqrt{K}\frac{\acute{\bx}_{\mathrm{R}}^{(k)}}{\left\lVert\acute{\bx}_{\mathrm{R}}^{(k)}\right\rVert}$
\State 8:~~~~~\textbf{end~If}
\State 9:~~\textbf{end While}
  \end{algorithmic}
\end{algorithm}
In Step 1, we initialize to
\begin{equation}
\acute{\bx}_{\mathrm{R}}^{(0)}=\sqrt{K}\frac{\widetilde{\bG}_{\mathrm{R}}^{\mathrm{T}}\mathbf{1}_{2N_{\mathrm{r}}}}{\left\lVert\widetilde{\bG}_{\mathrm{R}}^{\mathrm{T}}\mathbf{1}_{2N_{\mathrm{r}}}\right\rVert}\label{init}
\end{equation}
to resemble the maximum ratio estimate. The quantization using one-bit ADCs has been reflected in $\widetilde{\bG}_{\mathrm{R}}$ as shown in \eqref{sign_refine}, therefore, the maximum ratio estimate is based on the all-one vector. Steps 6 and 7 are based on projected gradient method \cite{Calamai:1987} to ensure the norm constraint. Note that a similar iterative algorithm was proposed in \cite{Wang:2014icc}. While our approach can support arbitrary constellations without any modification on Step 6 and 7, the constraint on the algorithm in \cite{Wang:2014icc} is separately adapted to each constellation, i.e., for $M$-QAM constellation, the constraint on each element of $\acute{\bx}_{\mathrm{R}}^{(k)}$ is set to $[-(\sqrt{M}-1)A,(\sqrt{M}-1)A]$ in \cite{Wang:2014icc} where $A$ is the average power normalization factor.

After obtaining the estimate $\check{\bx}_{\mathrm{R,ML}}^{(2)}$, the base station needs to perform normalization followed by symbol-by-symbol detection similar to the ZF-type detector in \eqref{zf_sym_detect}.  If we let $\bar{x}_{\mathrm{R,ML},k}$ be the $k$-th element of
\begin{equation}\label{norm_ml_est}
\bar{\bx}_{\mathrm{R,ML}}=\sqrt{K}\frac{\check{\bx}_{\mathrm{R,ML}}^{(2)}}{\left\lVert\check{\bx}_{\mathrm{R,ML}}^{(2)}\right\rVert},
\end{equation}
the nML symbol-by-symbol detection is
\begin{equation}\label{sym_by_sym}
  \hat{x}_{\mathrm{nML},k}=\argmin_{\acute{x}\in \cS}\left|(\bar{x}_{\mathrm{R,ML},k}+j\bar{x}_{\mathrm{R,ML},K+k})-\acute{x}\right|
\end{equation}
considering the fact that $\bar{\bx}_{\mathrm{R,ML}}$ is a $2K\times 1$ real vector.

\textbf{Remark 1:} As also discussed in \cite{Wang:2014icc}, the complexity of Algorithm~1 is dominated by the number of iterations because there are only two matrix-vector multiplications (one in the initialization in \eqref{init} and the other in the iteration loop in Algorithm 1) and numerical calculations to get $\triangledown f\left(\bz\right)_i$ in the iterative update.  The numerical calculations for updating $\triangledown f\left(\bz\right)_i$ in each iteration prevent performing the direct comparison with the ZF-type detector but it is reasonable to assume that the complexity of one iteration is less than that of the ZF-type detector that requires matrix inversion. As shown in Section \ref{simul}, the number of iterations for Algorithm 1 to converge can be moderate, e.g., less than 20.


\noindent \textbf{Remark 2:} Note that either nML detector based on \eqref{ml_est} or \eqref{ml_est_convex} is suboptimal compared to the original ML detector \eqref{ml_detect}.  The proposed nML detector is based on the ML estimation \eqref{ml_est_convex} over the $2K$-dimensional real space with the norm constraint. It can be the case that
\begin{align}
  &\sum_{i=1}^{2}\sum_{n=1}^{N_{\mathrm{r}}}\log\Phi\left(\sqrt{2\rho}\widetilde{\bg}_{\mathrm{R},n,i}^{\mathrm{T}}\check{\bx}^{(2)}_{\mathrm{R,ML}}\right)\\
  &\quad>\sum_{i=1}^{2}\sum_{n=1}^{N_{\mathrm{r}}}\log\Phi\left(\sqrt{2\rho}\widetilde{\bg}_{\mathrm{R},n,i}^{\mathrm{T}}\bx_{\mathrm{R}}\right)\label{xr_true}\\
  &\quad>\sum_{i=1}^{2}\sum_{n=1}^{N_{\mathrm{r}}}\log\Phi\left(\sqrt{2\rho}\widetilde{\bg}_{\mathrm{R},n,i}^{\mathrm{T}}\ddot{\bx}_{\mathrm{R}}\right)\label{xr_false}
\end{align}
with
\begin{equation}\label{diff}
  \left\lVert \bx_{\mathrm{R}}-\check{\bx}^{(2)}_{\mathrm{R,ML}}\right\rVert^2>\left\lVert \ddot{\bx}_{\mathrm{R}}-\check{\bx}^{(2)}_{\mathrm{R,ML}}\right\rVert^2
\end{equation}
where $\bx_{\mathrm{R}}$ is the true transmitted vector and $\ddot{\bx}_{\mathrm{R}}\in\mathcal{S}_{\mathrm{R}}^K\setminus \{\bx_{\mathrm{R}}\}$. In words, it can happen that the original ML detector makes the correct decision as in \eqref{xr_true} and \eqref{xr_false} while the estimated vector $\check{\bx}^{(2)}_{\mathrm{R,ML}}$ from \eqref{ml_est_convex} is closer to $\ddot{\bx}_{\mathrm{R}}$ that is not the true transmitted vector. The suboptimality, however, will not come into play when $N_{\mathrm{r}}$ is large as shown in \cite{Choi:2015}.


\subsection{Analyses in Asymptotic Regimes}

The estimates $\check{\bx}^{(1)}_{\mathrm{R,ML}}$ in \eqref{ml_est} and $\check{\bx}^{(2)}_{\mathrm{R,ML}}$ in \eqref{ml_est_convex} may not be the same in general because $\check{\bx}^{(1)}_{\mathrm{R,ML}}$ is selected from a circle $\|\acute{\bx}_{\mathrm{R}}\|^2=K$ while $\check{\bx}^{(2)}_{\mathrm{R,ML}}$ is selected from a ball $\|\acute{\bx}_{\mathrm{R}}\|^2\leq K$. Note that \eqref{ml_est} can have multiple local optimal solutions due to non-convexity of the constraint. If we let $\mathcal{X}^{(1)}$ be a set of all possible solutions of \eqref{ml_est}, the following lemma shows the relation between $\mathcal{X}^{(1)}$ and $\check{\bx}^{(2)}_{\mathrm{R,ML}}$ in the high SNR regime.

\begin{lemma}\label{same_proof}
When $\rho\rightarrow \infty$, we have
\begin{equation}
\check{\bx}^{(2)}_{\mathrm{R,ML}}\in \mathcal{X}^{(1)}.
\end{equation}
\end{lemma}
\begin{IEEEproof}
Please see Appendix \ref{App1}.
\end{IEEEproof}

Lemma \ref{same_proof} shows that the nML detector based on \eqref{ml_est_convex} will perform the same as in \eqref{ml_est} with the relaxed norm constraint in the high SNR regime.

It is necessary that different transmit vectors result in different quantized received signals to avoid possible detection errors because $\widetilde{\bg}_{\mathrm{R},n,i}$ in \eqref{sign_refine} is based on the sign refinement from the quantized received signal $\hat{y}_{\mathrm{R},n,i}$. Although it is difficult in general, we can explicitly derive the probability of which two different transmit vectors result in the same quantized received signal for a special case.

\textbf{Special case:} Consider two transmit vectors $\bx_1=\begin{bmatrix}x_1 & x_2 & \cdots & x_K\end{bmatrix}^{\mathrm{T}}$ and $\bx_2=\begin{bmatrix}-x_1 & x_2 & \cdots & x_K\end{bmatrix}^{\mathrm{T}}$ where $x_{k}$, which is selected from a standard $M$-ary constellation $\mathcal{S}$ with equal probability, is the transmit symbol of the $k$-th user. The two received signals are $\by_1=\sqrt{P}\bH\bx_1+\bn$ and $\by_2=\sqrt{P}\bH\bx_2+\bn$. Define $\hat{\by}_1$ and $\hat{\by}_2$ as the quantized outputs of $\by_1$ and $\by_2$ using one-bit ADCs. Assume each entry of $\bH$ follows IID Rayleigh fading and $\bn\sim \cC\cN(\mathbf{0}_{N_{\mathrm{r}}},\sigma^2\bI_{N_{\mathrm{r}}})$.

\begin{proposition}\label{prob_two_vec_proof}
For the special case,
\begin{equation}
  \mathrm{Pr}\left(\hat{\by}_1=\hat{\by}_2\right)=\left(\frac{2}{\pi}\arctan\sqrt{\frac{(K-1)P+\sigma^2}{P}}\right)^{2N_{\mathrm{r}}}.
\end{equation}
\end{proposition}
\begin{IEEEproof}
Please see Appendix \ref{App2}.
\end{IEEEproof}

The following corollary is a direct consequence of Proposition~\ref{prob_two_vec_proof}.
\begin{corollary}\label{prob_nr_infty}
For the special case, $\mathrm{Pr}\left(\hat{\by}_1=\hat{\by}_2\right)\rightarrow 0$ as $N_{\mathrm{r}}\rightarrow \infty$.
\end{corollary}
Corollary \ref{prob_nr_infty} shows that, for arbitrary SNR values, the two transmit vectors from the special case give different quantized signals as the number of receive antennas goes to infinity, which shows the benefit of massive MIMO with one-bit ADCs. Although it is hard to generalize Corollary \ref{prob_nr_infty} for arbitrary pair of transmit vectors, we expect a similar result would hold in general.

\subsection{Two-Stage nML Detector}\label{sec_nml_post}
To improve the performance of nML, we also propose a two-stage nML detector. The two-stage nML detector reduces the number of candidate transmit vectors, based on the output of the one-stage nML detector. Using the estimate $\bar{\bx}_{\mathrm{R,ML}}$ in \eqref{norm_ml_est} and the detected symbol $\hat{x}_{\mathrm{nML},k}$ in \eqref{sym_by_sym}, we define the candidate set of the $k$-th element
\begin{equation}\label{cand_form}
\mathcal{X}_k=\left\{\acute{x}\in \cS \left\rvert \frac{\left|(\bar{x}_{\mathrm{R,ML},k}+j\bar{x}_{\mathrm{R,ML},K+k})-\acute{x}\right|}{\left|(\bar{x}_{\mathrm{R,ML},k}+j\bar{x}_{\mathrm{R,ML},K+k})-\hat{x}_{\mathrm{nML},k}\right|}<c \right.\right\}
\end{equation}
where $c$ is a constant that controls the size of $\mathcal{X}_k$. With $\mathcal{X}_k$ for $k=1,\ldots,K$, we can define the reduced candidate set of the transmit vectors as
\begin{equation}\label{size_post}
\mathcal{X}=\left.\left\{\begin{bmatrix}\check{x}_1 & \cdots & \check{x}_K\end{bmatrix}^{\mathrm{T}} \right\rvert \check{x}_k\in \mathcal{X}_k,~\forall k\right\}.
\end{equation}
If we let
\begin{equation}
\mathcal{X}_{\mathrm{R}}=\left.\left\{\begin{bmatrix}\mathrm{Re}\left(\acute{\bx}\right)^{\mathrm{T}} & \mathrm{Im}\left(\acute{\bx}\right)^{\mathrm{T}}\end{bmatrix}^{\mathrm{T}} \right\rvert \acute{\bx}\in \mathcal{X}\right\},
\end{equation}
then the second stage of the two-stage nML detector can be defined as
\begin{align}\label{post_ml}
\hat{\bx}_{\mathrm{R,two-nML}}=\argmax_{\acute{\bx}_{\mathrm{R}}\in\mathcal{X}_{\mathrm{R}}} \prod_{i=1}^{2}\prod_{n=1}^{N_{\mathrm{r}}}\Phi\left(\sqrt{2\rho}\widetilde{\bg}_{\mathrm{R},n,i}^{\mathrm{T}}\acute{\bx}_{\mathrm{R}}\right).
\end{align}

The set $\mathcal{X}$ (or $\mathcal{X}_{\mathrm{R}}$) allows testing possible transmit vectors that are \textit{close} to the estimate $\bar{\bx}_{\mathrm{R,nML}}$. If we let the constant $c\rightarrow \infty$, the two-stage nML detector becomes the original ML detector. With a proper value of $c$, the numerical results in Section \ref{simul} show that it is possible to effectively improve the detection performance especially in the high SNR regime with marginal additional computational complexity. The concept of the two-stage nML detector is similar to that of sphere decoding that exploits reduced search space.

\subsection{Extension to Multicell Setting}\label{multicell_sec}

For the multicell scenario, the inter-cell interference should be taken into account for the proposed detectors.  It is reasonable to assume that the base station can accurately estimate the long-term statistic of the inter-cell interference. Without having any instantaneous CSI from out-of-cell users, however, the base station will consider the inter-cell interference as additional AWGN, i.e.,
\begin{equation}\label{multicell_assum}
\sqrt{P}\sum_{\stackrel{m=1}{m\neq i}}^{N_{\mathrm{c}}}\sum_{k=1}^{K}\bh_{i,mk} x_{mk}\sim \cC\cN(\mathbf{0}_{N_{\mathrm{r}}},P\eta_{i}^2\bI_{N_{\mathrm{r}}})
\end{equation}
where $\eta_i$ captures the long-term inter-cell interference statistic, e.g., pathloss and shadowing. Then the effective signal-to-interference-noise ratio (SINR) at the $i$-th base station is
\begin{equation}
  \rho_{i,\mathrm{MC}}=\frac{P}{P\eta_i^2+\sigma^2},
\end{equation}
and the proposed detectors can be adapted to the multicell scenario by substituting $\rho$ in \eqref{ml_est_convex} and \eqref{post_ml} with $\rho_{i,\mathrm{MC}}$.  

\section{Channel Estimation with One-Bit ADCs}\label{sec_channel_est}
For a coherent detector, the CSI is normally obtained through an estimate of the channel.  In this section, we develop channel estimation techniques for one-bit ADCs at the base station.  We focus on estimating $\bg_n$, i.e., the channel between the receive antenna $n$ and $K$ users, instead of $\bh_k$.

We consider a block fading channel to develop channel estimation techniques.  We assume the channel is static for $L$ channel uses in a given fading block and changes independently from block-to-block.  The received signal at the $n$-th antenna for the $\ell$-th channel use in the $m$-th fading block is given as
\begin{equation}
  y_{n,m}[\ell] = \sqrt{\rho}\bg_{n,m}^{*}\bx_m[\ell]+w_{n,m}[\ell].
\end{equation}

Let the first $T<L$ channel uses be devoted for a training phase and the remaining $L-T$ channel uses be dedicated to a data communication phase.  Put the first $T$ received signals during the training phase into a vector form as
\begin{equation}
  \by_{n,m,\mathrm{train}} = \sqrt{\rho}\bX_{m,\mathrm{train}}^{*}\bg_{n,m}+\bw_{n,m,\mathrm{train}}
\end{equation}
where
\begin{align}
    \by_{n,m,\mathrm{train}}&=\begin{bmatrix}y_{n,m}[0] & \cdots & y_{n,m}[T-1]\end{bmatrix}^{*}\in \mathbb{C}^{T \times 1},\\
    \bX_{m,\mathrm{train}}&=\begin{bmatrix}\bx_m[0] & \cdots & \bx_m[T-1]\end{bmatrix}\in\mathbb{C}^{K \times T},\\
    \bw_{n,m,\mathrm{train}}&=\begin{bmatrix}w_{n,m}[0] & \cdots & w_{n,m}[T-1]\end{bmatrix}^{*}\in\mathbb{C}^{T \times 1}.
\end{align}

In the training phase, $\bX_{m,\mathrm{train}}$ is known to the base station but $\bg_{n,m}$ must be estimated.  While arbitrary training matrices are possible, for simulation purpose in Section \ref{simul}, we focus on unitary training where $\bX_{m,\mathrm{train}}$ satisfies \cite{Santipach:2010}
\begin{align}
\bX_{m,\mathrm{train}} \bX_{m,\mathrm{train}}^{*} =T\bI_{K}\quad &\text{if }K<T.
\end{align}
The normalization term $T$ ensures the average transmit power equals to $P$ in each channel use.

Similar to the previous sections, we reformulate all expressions into the real domain as
\begin{equation}\label{train_phase}
  \by_{\mathrm{R},n,m,\mathrm{train}} = \sqrt{\rho}\bX_{\mathrm{R},m,\mathrm{train}}^{\mathrm{T}}\bg_{\mathrm{R},n,m}+\bw_{\mathrm{R},n,m,\mathrm{train}}
\end{equation}
where
\begin{align}
\by_{\mathrm{R},n,m,\mathrm{train}}&=\begin{bmatrix}\mathrm{Re}\left(\by_{n,m,\mathrm{train}}\right) \\ \mathrm{Im}\left(\by_{n,m,\mathrm{train}}\right)\end{bmatrix}\in\mathbb{R}^{2T\times 1},\\
\bX_{\mathrm{R},m,\mathrm{train}}&=\begin{bmatrix} \mathrm{Re}(\bX_{m,\mathrm{train}}) & -\mathrm{Im}(\bX_{m,\mathrm{train}}) \\ \mathrm{Im}(\bX_{m,\mathrm{train}}) & \mathrm{Re}(\bX_{m,\mathrm{train}})\end{bmatrix}\in\mathbb{R}^{2K\times 2T},\\
\bg_{\mathrm{R},n,m}&=\begin{bmatrix}\mathrm{Re}\left(\bg_{n,m}\right) \\ \mathrm{Im}\left(\bg_{n,m}\right)\end{bmatrix}\in\mathbb{R}^{2K\times 1},\\
\bw_{\mathrm{R},n,m,\mathrm{train}}&=\begin{bmatrix}\mathrm{Re}\left(\bw_{n,m,\mathrm{train}}\right) \\ \mathrm{Im}\left(\bw_{n,m,\mathrm{train}}\right)\end{bmatrix}\in\mathbb{R}^{2T\times 1}.
\end{align}
It is important to point out that \eqref{train_phase} has the same form as \eqref{data_phase} while the roles of the channel and the transmitted signal are reversed.  Therefore, using the same techniques that we exploited for the detectors, we can develop channel estimators based on the one-bit ADC outputs and $\bX_{\mathrm{R},m,\mathrm{train}}$.

We define the $i$-th column of $\bX_{\mathrm{R},m,\mathrm{train}}$ as $\bx_{\mathrm{R},m,train,i}$ and the $i$-th output of the one-bit ADC as
\begin{equation}
\hat{y}_{\mathrm{R},n,m,train,i} = \mathrm{sgn}(y_{\mathrm{R},n,m,train,i})
\end{equation}
where $y_{\mathrm{R},n,m,train,i}$ is the $i$-th element of $\by_{\mathrm{R},n,m,\mathrm{train}}$.  Note that there are $2T$ one-bit ADC outputs in total for the $n$-th receive antenna, i.e., $T$ outputs for each of the real and imaginary parts of the received signal.  Based on $\hat{y}_{\mathrm{R},k,m,train,i}$, the base station performs the sign-refinement as
\begin{equation}
\widetilde{\bx}_{\mathrm{R},m,train,i}=\hat{y}_{\mathrm{R},n,m,train,i}\bx_{\mathrm{R},m,train,i},
\end{equation}
and the ML channel estimator is given as
\begin{align}
 \check{\bg}_{\mathrm{R},n,m,\mathrm{ML}}&=\argmax_{\acute{\bg}_{\mathrm{R}}\in\mathbb{R}^{2K\times 1}} \prod_{i=1}^{2T}\Phi\left(\sqrt{2\rho}\widetilde{\bx}_{\mathrm{R},m,train,i}^{\mathrm{T}}\acute{\bg}_{\mathrm{R}}\right)\\
&=\argmax_{\acute{\bg}_{\mathrm{R}}\in\mathbb{R}^{2K\times 1}} \sum_{i=1}^{2T}\log\left(\Phi\left(\sqrt{2\rho}\widetilde{\bx}_{\mathrm{R},m,train,i}^{\mathrm{T}}\acute{\bg}_{\mathrm{R}}\right)\right).\label{ml_ch_est}
\end{align}

Because $\Phi(\cdot)$ is a log-concave function, and there is no constraint on $\acute{\bg}_{\mathrm{R}}$, it is possible to solve \eqref{ml_ch_est} using standard convex optimization methods (or suitably modified Algorithm 1).  Unfortunately, this channel estimator tends to overestimate the norm
\begin{align}
  \left\lVert \check{\bg}_{\mathrm{R},n,m,\mathrm{ML}} \right\rVert > \left\lVert \bg_{\mathrm{R},n,m} \right\rVert
\end{align}
due to the fact that $\log \Phi(\cdot)$ is an increasing function where $\bg_{\mathrm{R},n,m}$ is the true channel vector.

To overcome this problem, we impose a norm constraint on $\acute{\bg}_{\mathrm{R}}$ and convert \eqref{ml_ch_est} to
\begin{align}
\check{\bg}_{\mathrm{R},n,m,\mathrm{ML}}=\argmax_{\substack{\acute{\bg}_{\mathrm{R}}\in\mathbb{R}^{2K \times 1}\\ \|\acute{\bg}_{\mathrm{R}}\|^2\leq K}} \sum_{i=1}^{2T}\log\left(\Phi\left(\sqrt{2\rho}\widetilde{\bx}_{\mathrm{R},m,train,i}^{\mathrm{T}}\acute{\bg}_{\mathrm{R}}\right)\right)\label{ml_ch_est_norm_const}
\end{align}
using the fact that $\mathbb{E}\left[\left\lVert \bg_{\mathrm{R},n,m} \right\rVert^2\right]=K$ for most of channel models.  The norm constraint can be further optimized if the base station knows the long-term statistic of the channel norm.

For comparison purposes, we also define a simple ZF-type channel estimator as
\begin{equation}\label{zf_ch_est}
\check{\bg}_{\mathrm{R},n,m,\mathrm{ZF}}=\sqrt{K}\frac{\left(\bX_{\mathrm{R},m,\mathrm{train}}^{\mathrm{T}}\right)^{\dagger}\hat{\by}_{\mathrm{R},n,m,\mathrm{train}}}{\left\lVert\left(\bX_{\mathrm{R},m,\mathrm{train}}^{\mathrm{T}}\right)^{\dagger}\hat{\by}_{\mathrm{R},n,m,\mathrm{train}}\right\rVert}
\end{equation}
which is forced to satisfy $\|\check{\bg}_{\mathrm{R},n,m,\mathrm{ZF}}\|^2=K$.  We numerically compare our ML channel estimator and the ZF-type channel estimator in Section \ref{simul}.

\section{Simulation Results}\label{simul}
\begin{figure}[t]
  \centering
  \includegraphics[width=1\columnwidth]{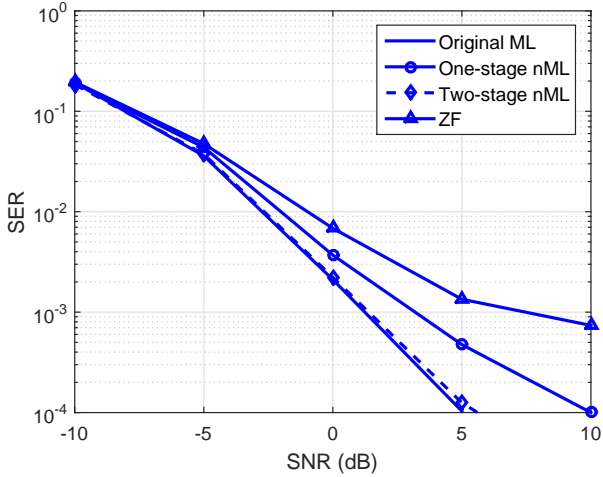}\\
  \caption{SER vs. SNR (in dB) of four detectors with $N_{\mathrm{r}}=32$, $K=4$, and $M=4$ (QPSK).  The original ML, the one-stage nML, the two-stage nML, and the ZF-type detectors are compared.}\label{K4M4N32}
\end{figure}
\begin{figure}
\centering
\includegraphics[width=1\columnwidth]{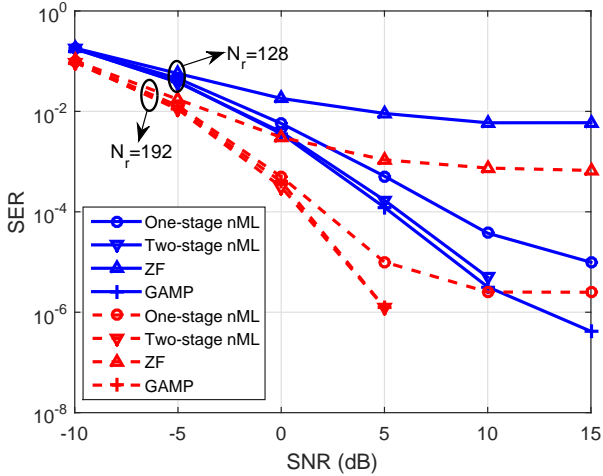}
\caption{SER vs. SNR in dB scale with $M=8$ (8PSK), $K=8$, and different values of $N_{\mathrm{r}}$.  The GAMP detector is from \cite{Wang:2015icc}.}\label{K8M8}
\end{figure}
\begin{figure}
\centering
\includegraphics[width=1\columnwidth]{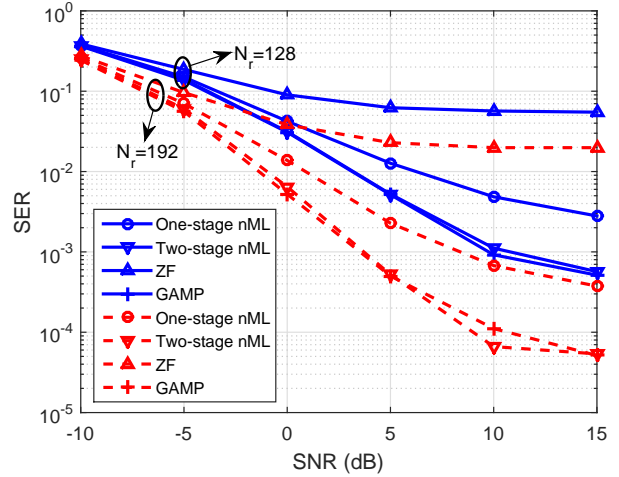}
\caption{SER vs. SNR in dB scale with $M=16$ (16QAM), $K=8$, and different values of $N_{\mathrm{r}}$. The GAMP detector is from \cite{Wang:2015icc}.}\label{K8M16}
\end{figure}


We perform Monte-Carlo simulations to evaluate the proposed techniques.  The one-stage nML detector is based on Algorithm 1 with a termination threshold $\epsilon=10^{-3}$ and step size $\kappa=0.01$. For the two-stage nML detector, we let the constant $c=1.3$ in \eqref{cand_form}. We first consider the single cell scenario to compare detector performance. Then we take the multicell scenario into account.
\subsection{Single Cell Scenario}
We assume IID Rayleigh fading channels, i.e., all elements of $\bH$ are distributed as $\cC\cN(0,1)$, although the distribution of the channel was not explicitly incorporated into the proposed detector designs. For the time being, we assume the base station has perfect CSI and evaluate the detectors. Later, we evaluate the detectors with imperfect CSI. We use the average symbol error rate (SER) which is defined as
\begin{equation}
  \mathrm{SER} = \frac{1}{K}\sum_{n=1}^{K}\mathbb{E}\left[\mathrm{Pr}\left(\hat{x}_n\neq x_n\mid \bx~\text{sent},\bH,\bn,\rho,K,N_{\mathrm{r}},\cS \right)\right]
\end{equation}
for the performance metric where the expectation is taken over $\bx$, $\bH$, and $\bn$.
\begin{figure*}
\centering
\subfloat[$\rho=0$dB.]{
\includegraphics[width=1\columnwidth]{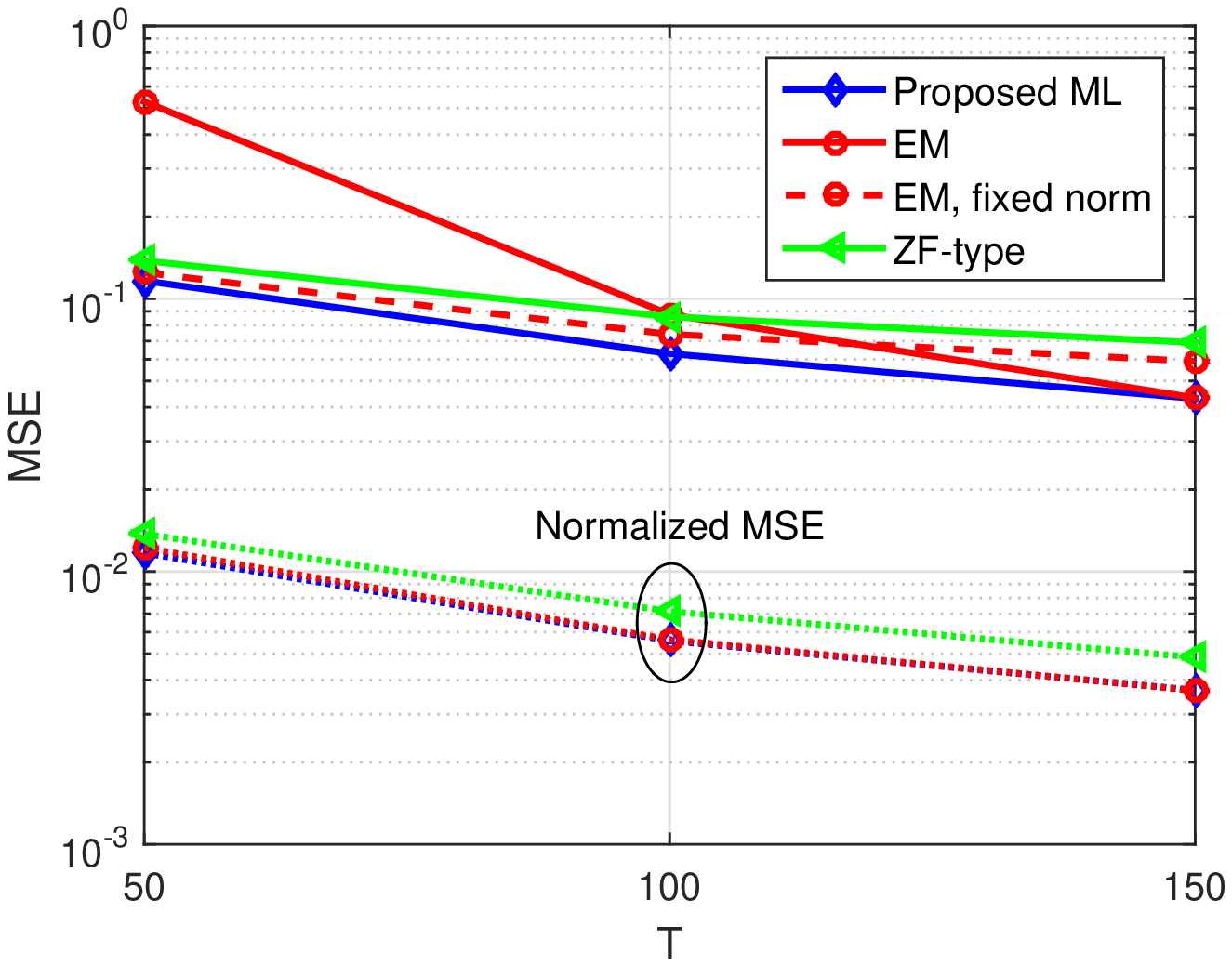}
\label{mse1}
}
\subfloat[$\rho=20$dB.]{
\includegraphics[width=1\columnwidth]{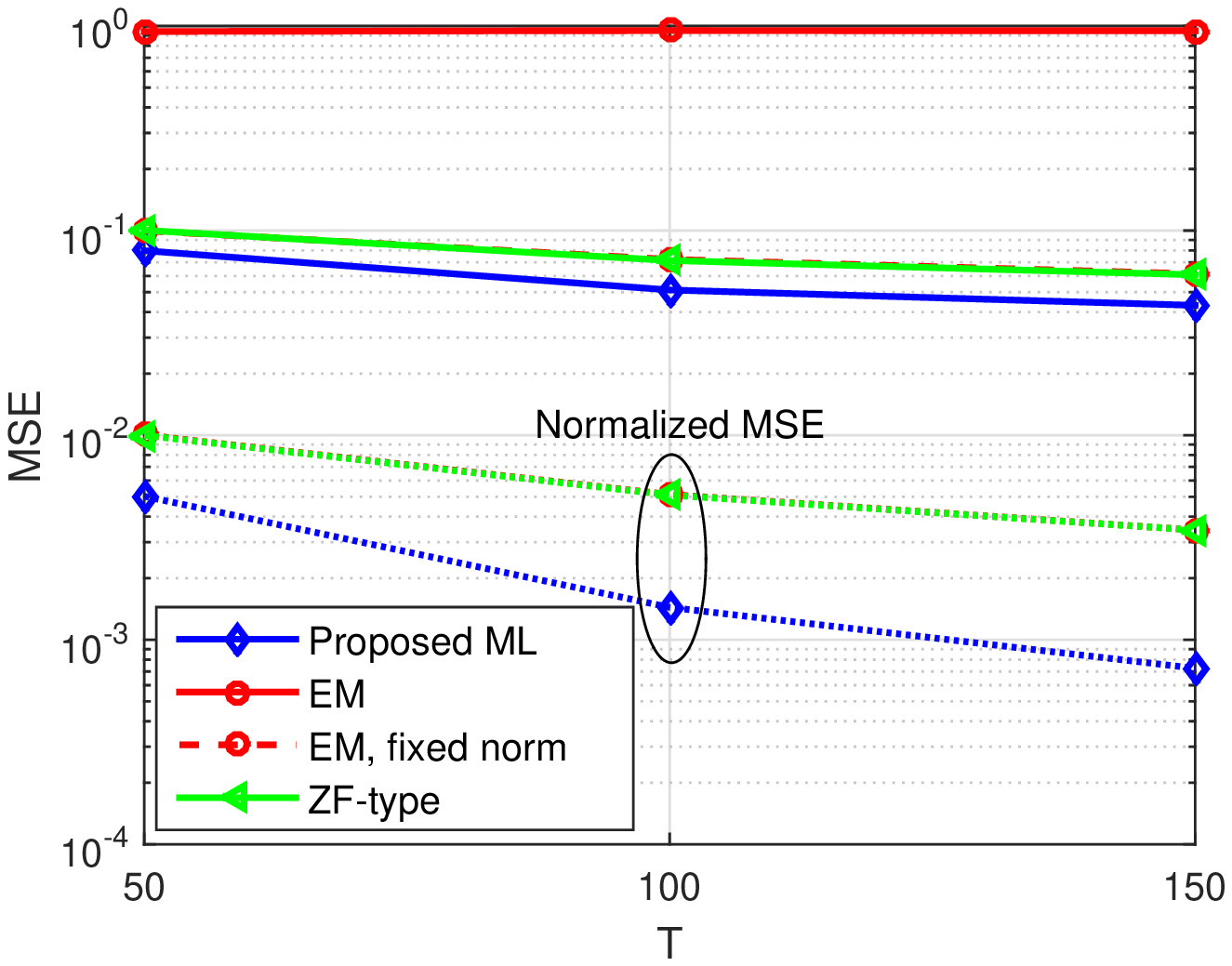}
\label{mse2}
}
\caption{MSE and normalized MSE of different channel estimators with $K=8$ and different values of $T$ and SNR .  Expectation-maximization (EM) estimation is from \cite{Ivrlac:2007}, and ``EM, fixed norm'' is the same as EM except the norm square fixed to $K$.}\label{ch_est_fig}
\label{mse_fig}
\end{figure*}


We first compare the four detectors: 1) the original ML detector \eqref{ml_detect} that is based on exhaustive search over all possible transmitted vectors, 2) the one-stage nML detector \eqref{ml_est_convex} that is based on convex optimization, 3) the two-stage nML detector explained in Section \ref{sec_nml_post}, and 4) the ZF-type detector.  Due to the computational complexity of the original ML detector, we set $K=4$, $M=4$ (QPSK) for all users, and $N_{\mathrm{r}}=32$ which may not be considered as massive MIMO.  We plot SERs of the four detectors in Fig. \ref{K4M4N32}. The figure shows that the one-stage nML detector is suboptimal compared to the original ML detector as discussed in Remark~2 of Section~\ref{convex_ml} while the two-stage nML detector gives almost the same performance with the original ML detector. Further, both the nML detectors outperform the ZF-type detector.

In Fig. \ref{K8M8}, we plot the SERs of the one- and two-stage nML detectors, the ZF-type detector, and the detector based on the GAMP algorithm from \cite{Wang:2015icc} according to SNR with $K=8$, $M=8$ (8PSK), and different values of $N_{\mathrm{r}}$. We have modified the GAMP algorithm to have adaptive step sizes as in \cite{Vila:2015} for better performance. We do not consider the original ML detector in this scenario because of its excessive complexity, i.e., the detector needs to compare $8^8=2^{24}$ possible transmit vectors. The figure shows that the two-stage nML detector  and GAMP detector are comparable and outperform the one-stage nML and ZF-type detectors with the same number of $N_{\mathrm{r}}$.  The ZF-type detector suffers from an error rate floor while other detectors do not have such floor until $10^{-5}$ SER. With small SERs, it is possible to use weaker channel coding (with higher code rate or shorter block length) to improve the system throughput, which has been shown using binary symmetric channels in \cite{polyanskiy2010channel}.

We adopt the same system setup with Fig. \ref{K8M8} except a $M=16$ (16QAM) constellation for data symbol in Fig. \ref{K8M16}.  The two-stage nML detector and GAMP detector are still comparable; however, the two detectors also suffer from an error rate floor. With $N_{\mathrm{r}}=196$, the error rate floors are mitigated for all detectors, which shows the benefit of massive MIMO for using one-bit ADCs.

Figs. \ref{K8M8} and \ref{K8M16} both show that the proposed detectors give much better SER performance compared to the ZF-type detector with the same number of receive antennas, which means that the proposed detectors are able to support more users than the ZF-type detector with the same system setup. Therefore, if the base station has sufficient computational power, it is always beneficial to use the proposed detectors than the ZF-type detector.  Note that the computational power at the base station is related to the digital baseband processing in Fig. \ref{concept_fig} and different from having one-bit ADCs. Therefore, the benefit of using one-bit ADCs, e.g., power consumption and cost, still holds for the proposed detectors although higher computational power is required for the digital baseband processing at the base station.

\begin{table}[t] \centering
\caption{Average number of iterations for Algorithm 1 and size of $\mathcal{X}$ in \eqref{size_post} with $K=8$.}\label{avg_iter_num}
\begin{tabular}{|c||c|c|}
  \hline
  & Algorithm 1 & Size of $\mathcal{X}$ \\
  \hline
  $M=8$, $N_{\mathrm{r}}=128$ & 18.3721 & 2.6475 \\
  \hline
  $M=8$, $N_{\mathrm{r}}=192$ & 12.1172 & 1.9522 \\
  \hline
  $M=16$, $N_{\mathrm{r}}=128$ & 18.8378 & 4.7196 \\
  \hline
  $M=16$, $N_{\mathrm{r}}=192$ & 13.0174 & 4.6471 \\
  \hline
\end{tabular}
\end{table}

To evaluate the complexity of the proposed detectors, we compare the average number of iterations for Algorithm 1 to converge for the one-stage nML detector and the average size of $\mathcal{X}$ in \eqref{size_post} for the two-stage nML detector, with $K=8$ and different values of $M$ and $N_{\mathrm{r}}$ in Table \ref{avg_iter_num}. It shows that Algorithm 1 requires less than 20 iterations to converge in average and the additional comparison, i.e., the size of $\mathcal{X}$, performed in the two-stage nML detector is marginal. It is interesting that the iteration number of Algorithm 1 decreases with the number of antennas, which shows that the proposed detectors based on Algorithm 1 will become more efficient with large $N_{\mathrm{r}}$.

Now, we evaluate the ML and ZF-type channel estimators discussed in Section \ref{sec_channel_est}.  We focus on estimating $\bg_{n,m}$, i.e., the channel between the $n$-th receive antenna and $K$ users.  We define the mean squared error (MSE) of a channel estimator x as
\begin{equation}
  \mathrm{MSE}_{\mathrm{x}}=\frac{1}{K}\mathbb{E}\left[\left\lVert\bg_{n,m}-\check{\bg}_{n,m,x}\right\rVert^2\right],
\end{equation}
and the normalized MSE (NMSE) as
\begin{equation}
  \mathrm{NMSE}_{\mathrm{x}}=\frac{1}{K}\mathbb{E}\left[\left\lVert\frac{\bg_{n,m}}{\|\bg_{n,m}\|}-\frac{\check{\bg}_{n,m,x}}{\|\check{\bg}_{n,m,x}\|}\right\rVert^2\right],
\end{equation}
which are used as performance metrics.  The expectations are taken over $\bg_{n,m}$.  In Fig. \ref{ch_est_fig}, we compare the proposed ML channel estimator to the ZF-type estimator and the expectation-maximization (EM) method from \cite{Ivrlac:2007} with different training lengths $T$ and SNR values with $K=8$.  Regarding MSE, the proposed ML estimator outperforms other estimators for both $\rho=0$ and 20dB cases. The EM method performs well only when $T$ is large with $\rho=0$dB and fails to estimate the channel norm when $\rho=20$dB as also shown in \cite{Ivrlac:2007}. The EM method with the norm fixed to $K$ gives better performance than the ZF-type estimator when $\rho=0$dB while the two estimators become exactly the same with $\rho=20$dB. Regarding NMSE, the proposed ML estimator and the EM method (and the one with the fixed norm as well) are comparable when $\rho=0$dB while the ML estimator outperforms the EM method with $\rho=20$dB.

\begin{figure}
\centering
\includegraphics[width=1\columnwidth]{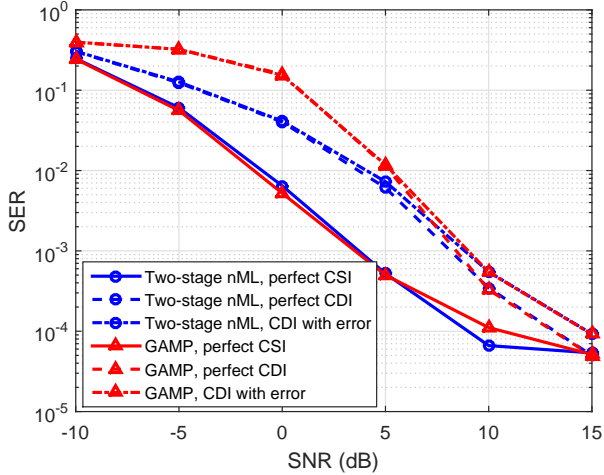}
\caption{SER vs. SNR in dB scale with $M=16$ (16QAM), $K=8$, and $N_{\mathrm{r}}=192$.  The two-stage nML detector and the GAMP detector from \cite{Wang:2015icc} are compared with different assumptions on channel information.}\label{ch_est_SER}
\end{figure}

To verify the effect of channel estimation, we plot the SERs of the two-stage nML detector  and GAMP detector with different assumptions on CSI; perfect CSI, perfect CDI, and CDI with error that is defined as
\begin{equation}
\frac{\bg_{n,m}/\|\bg_{n,m}\|+\sqrt{\mathrm{NMSE}}\bee_{n,m}/\|\bee_{n,m}\|}{\left\lVert\bg_{n,m}/\|\bg_{n,m}\|+\sqrt{\mathrm{NMSE}}\bee_{n,m}/\|\bee_{n,m}\|\right\rVert}
\end{equation}
where the element of $\bee_{n,m}$ is distributed as $\cC\cN(0,1)$. We set $K=8$, $M=16$, $N_{\mathrm{r}}=192$, and $\mathrm{NMSE}=10^{-2}$. It is shown in Fig. \ref{ch_est_SER} that the two-stage nML detector outperforms the GAMP detector using CDIs when SNR is low, which shows that the GAMP detector requires more accurate channel norm information than the proposed detector for low SNR values.

\begin{figure}
\centering
\includegraphics[width=1\columnwidth]{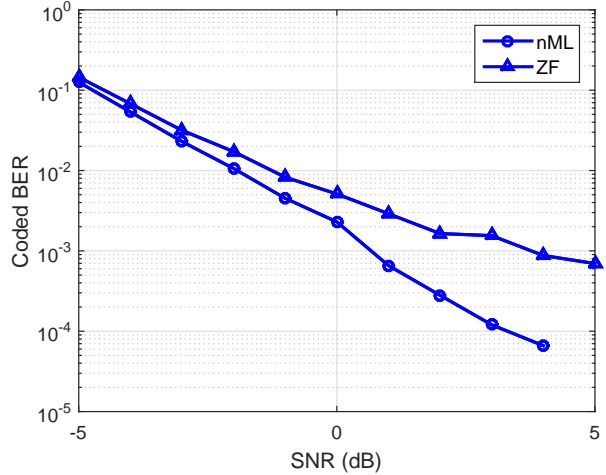}
\caption{Coded BER vs. SNR in dB scale with $M=8$ (8PSK), $K=4$, and $N_{\mathrm{r}}=64$. The proposed nML and ZF-type detectors are compared using the rate 1/2 LDPC code adopted in the IEEE 802.11ad standard.}\label{ldpc_plot}
\end{figure}

To compare the detectors in a practical setting, we combine the detectors with a low-density-parity-check (LDPC) code.  We assume the base station has perfect CSI for this study.  We adopt a rate 1/2 LDPC code with the block length of 672 bits from the IEEE 802.11ad standard \cite{11ad}.  After hard detection by the detectors, the estimated symbols (or bits) are decoded using the bit-flipping decoding algorithm \cite{Rao:2015}.  The coded bit error rates (BERs) of the one-stage nML detector and ZF-type detector according to SNR with $K=4$, $N=64$, and $M=8$ are shown in Fig. \ref{ldpc_plot}. The figure clearly shows that the nML detector outperforms the ZF-type detector even for this practical setting.  Further improvements could be expected if further work is put into deriving an appropriate soft decision decoding metric, which requires the probability distribution of $\check{\bx}^{(2)}_{\mathrm{R,ML}}$ in \eqref{ml_est_convex}. It is also possible to exploit the approximated soft metric as in \cite{Studer:2011}, but we leave this for future work.


\subsection{Multicell Scenario}

For the last numerical study, we consider the multicell scenario where the detailed simulation parameters are listed in Table \ref{table1}.  We consider two different user dropping scenarios. 1) All users except a typical user in the center cell are randomly dropped within corresponding cells.  The typical user is located with the distance $d$ (and a random angle per drop) from the base station in the center cell.  2) All users except the users in the center cell are randomly dropped within corresponding cells.  The users in the center cell are randomly dropped within the range $(d-20\mathrm{m}, d+20\mathrm{m})$ from the center cell base station.  The second scenario can be considered as \textit{coordinated} uplink transmission with user scheduling that selects users with similar received signal power while the first scenario corresponds to \textit{uncoordinated} uplink transmission.  We consider coded BER of the typical user for the first scenario while BERs of all $K$ users in the center cell are averaged for the second scenario. We neglect shadowing effect and $\eta_i$ in \eqref{multicell_assum} is based on the distances between the center cell base station and out-of-cell users.

\begin{figure}
\centering
\includegraphics[width=1\columnwidth]{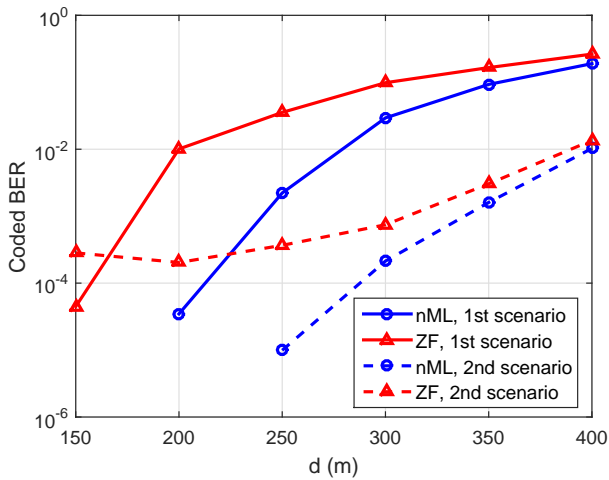}
\caption{Coded BER vs. $d$ (in meters) for the multicell setting with parameters in Table \ref{table1}.  Coordinated and uncoordinated user dropping scenarios are compared for the nML and ZF-type detectors.}\label{multicell}
\end{figure}

The BER results for these two scenarios according the distance $d$ are plotted in Fig. \ref{multicell}.  We can see that the one-stage nML detector outperforms the ZF-type detector for both user dropping scenarios.  As $d$ increases, the BER performance of both detectors becomes worse because of the reduced received signal power. Note that the BER performance of the second scenario is much better than that of the first scenario when $d$ is large.  For large $d$, the received signal power of the typical user in the first scenario is overwhelmed by other users' received signals (the near-far effect), resulting in poor BER performance.  If all users experience similar SINR as in the second scenario, however, the BER performance is quite good even with large $d$.  This shows that the nML detector will perform well with proper user scheduling or uplink power control, which are already common in current cellular systems.  Note that the ZF-type detector suffers from the notable error rate floor for the second scenario when $d$ is small (that corresponds to the high SNR regime for the single cell scenario) while the proposed nML detector does not have such floor until $10^{-5}$ BER.  In the first scenario, there is no error rate floor even for the ZF-type detector because the received signal of the typical user overwhelms other users' received signals.

\begin{table} \centering
\caption{Multicell Simulation Parameters}\label{table1}
~\\
\begin{tabular}{c c}
\toprule[.2em]
{\textrm{Parameter}} & \multicolumn{1}{c}{\textrm{Assumption}}\\
\midrule[.1em]
{\textrm{Cell layout}} & {\textrm{57 hexagonal cells}}\\
{\textrm{Cell radius}} & {500m}\\
{\textrm{\# of RX antennas per BS ($N_{\mathrm{r}}$)}} & {64}\\
{\textrm{\# of TX antennas per user}} & {1}\\
{\textrm{\# of users per cell ($K$)}} & {4}\\
{\textrm{Min. dist. btw. BS and user}} & {100m}\\
{\textrm{User transmit power}} & {23dBm}\\
{\textrm{Path loss per km}} & {131.1+42.8$\log_{10}(\mathrm{dist.})$ dB}\\
{\textrm{System bandwidth}} & {5MHz}\\
{\textrm{Noise spectral density}} & {-174dBm/Hz}\\
{\textrm{Noise figure}} & {5dB}\\
{\textrm{Constellation}} & {8PSK}\\
{\textrm{Channel coding}} & {Rate 1/2 LDPC from 802.11ad}\\
\bottomrule[.2em]
\end{tabular}
\end{table}

\section{Conclusion}\label{conclusion}
We proposed a unified framework of detection and channel estimation techniques for uplink multiuser massive MIMO systems using a pair of one-bit ADCs at each antenna. The proposed techniques are based on off-the-shelf convex optimization methods, which makes it easy to implement in practice. We proposed two nML detectors, i.e., one-stage and two-stage nML detectors, which give better performance than the ZF-type detector for all range of SNR regimes and number of antennas. The two-stage nML detector reduces the number of candidate transmit vectors using the output of the one-stage nML detector and exploits the ML detector structure to further improve detection performance of the one-stage nML detector. Numerical studies showed that the proposed detectors are able to perform well even with not-so-large number of antennas, robust to inaccurate channel estimation, and outperform the ZF-type detector for practical channel coded and multicell settings. We also proposed a ML channel estimator that can effectively estimate not only the direction but also the norm of the channel even with one-bit ADCs. Because of the unified structure of the proposed detectors and channel estimator, same hardware can be used for both tasks, which make the proposed techniques more attractive for uplink massive MIMO systems using one-bit ADCs.

There are several directions for future research. The proposed detectors can be extended to frequency selective channel considering their possible use in mmWave systems, for example by extending the results in \cite{Wang:2015TWC,Studer:2015}. To make the proposed ML channel estimator more practical, the training overhead should be reduced. As in \cite{Choi:2014JSTSP,Song:2014JSTSP}, it may be possible for the proposed ML channel estimator to exploit channel statistics, e.g., the temporal and spatial correlation, to reduce the training overhead. It would be also interesting to consider maximum a posteriori (MAP) detector and channel estimator using one-bit ADCs and compare them with the proposed techniques where MAP detector and channel estimator with low-resolution ADCs have been studied in \cite{Studer:2015}.

\appendices
\section{Proof of Lemma \ref{same_proof}}\label{App1}

Recall the estimator in \eqref{ml_est_convex}. When $\rho \rightarrow \infty$,
\begin{equation}
\left|\sqrt{2\rho}\widetilde{\bg}^{\mathrm{T}}_{\mathrm{R},n,i}\acute{\bx}_{\mathrm{R}}\right|\rightarrow \infty
\end{equation}
unless $\widetilde{\bg}^{\mathrm{T}}_{\mathrm{R},n,i}\acute{\bx}_{\mathrm{R}}= 0$. Therefore, the estimator \eqref{ml_est_convex} finds $\check{\bx}_{\mathrm{R,ML}}^{(2)}$ that satisfies
\begin{equation}
  \widetilde{\bg}^{\mathrm{T}}_{\mathrm{R},n,i}\check{\bx}^{(2)}_{\mathrm{R,ML}}> 0
\end{equation}
for all $1\leq n\leq N_{\mathrm{r}}$ and $1\leq i\leq 2$ in the high SNR regime because $\Phi(t)$ is an increasing function of $t$ but upper bounded by 1. Then,
\begin{align}
  &\sum_{i=1}^{2}\sum_{n=1}^{N_{\mathrm{r}}}\log\Phi\left(\sqrt{2\rho}\widetilde{\bg}_{\mathrm{R},n,i}^{\mathrm{T}}\check{\bx}^{(2)}_{\mathrm{R,ML}}\right)\\
  &\quad >\sum_{i=1}^{2}\sum_{n=1}^{N_{\mathrm{r}}}\log\Phi\left(\sqrt{2\rho}\widetilde{\bg}_{\mathrm{R},n,i}^{\mathrm{T}}\alpha \check{\bx}^{(2)}_{\mathrm{R,ML}}\right)
\end{align}
with arbitrary $0<\alpha<1$.  Therefore, the norm square of $\check{\bx}^{(2)}_{\mathrm{R,ML}}$ always becomes $K$ due to the norm constraint, and
\begin{equation}
  \check{\bx}^{(2)}_{\mathrm{R,ML}}\in \mathcal{X}^{(1)}
\end{equation}
which finishes the proof.

\section{Proof of Proposition \ref{prob_two_vec_proof}}\label{App2}
First, we show that if
\begin{align}
u \sim \mathcal{N}(0, \sigma_{u}^2),~v\sim \mathcal{N}(0, \sigma_{v}^2),
\end{align}
we have
\begin{align}
	&\mathrm{Pr}\left(\mathrm{sgn}(u - v)  =  \mathrm{sgn}(u + v)\right)\\
	&=\int_{ - \infty }^{ + \infty } {\frac{1}{{\sqrt {2 \pi \sigma_u^2 } }}} {e^{ - \frac{{{u^2}}}{{2 \sigma_{u}^2 }}}}\int_{ - \left| u \right|}^{\left| u \right|} {\frac{1}{{\sqrt{2 \pi \sigma_v^2 }  }}{e^{ - \frac{v^2}{2 \sigma_{v}^2}}}dv} du\\
	&=\frac{1}{{\pi  \sigma_u \sigma_v }} \int_{ - \infty }^{ + \infty }  {e^{ - \frac{{{u^2}}}{{2 \sigma_{u}^2}}}}\int_{0}^{\left| u \right|} {{e^{ - \frac{v^2}{2 \sigma_v^2}}}dv} du\\
	&=\frac{2}{{\pi  \sigma_u \sigma_v }} \int_{ 0 }^{ + \infty } {{e^{ - \frac{{{u^2}}}{2 \sigma_u^2 }}}} \int_0^{ u } {{e^{ - \frac{v^2}{2 \sigma_v^2}}}dv} du \\
	&\stackrel{(a)}{=} \frac{2}{\pi \sigma_v^2} \int_0^{+ \infty} e^{-\frac{w^2}{2 \sigma_v^2}} \int_0^{ \frac{\sigma_u}{\sigma_v} w} e^{-\frac{v^2}{\sigma_v^2}} dv dw \\
	&=\frac{2}{\pi }\arctan \frac{\sigma_u}{\sigma_v},
\end{align}
where (a) follows by letting $u \triangleq w\frac{\sigma_u}{\sigma_v} $.

To prove Proposition \ref{prob_two_vec_proof}, note that
\begin{align}
    \sqrt{P}\bH \bx_1+\bn &= \sqrt{P}\bh_1 x_1 + \sqrt{P}\bH' \bx'+\bn,\\
    \sqrt{P}\bH \bx_2+\bn &= -\sqrt{P}\bh_1 x_1 + \sqrt{P}\bH' \bx'+\bn
\end{align}
where
\begin{align}
    \bH'&= \begin{bmatrix}\bh_2 & \bh_3 \cdots \bh_{K}\end{bmatrix},\\
    \bx'&= \begin{bmatrix}x_2 & x_3 & \cdots & x_{K}\end{bmatrix}^{\mathrm{T}}.
\end{align}
Because of the assumptions on $\bH$ and $x_k$, we have
\begin{align}
  \sqrt{P}\bh_1 x_1 &\sim \mathcal{CN}(\mathbf{0}_{N_{\mathrm{r}}}, P\bI_{N_{\mathrm{r}}}),\\
  \sqrt{P}\bH' \bx'+\bn &\sim \mathcal{CN}(\mathbf{0}_{N_{\mathrm{r}}}, ((K-1)P+\sigma^2)\bI_{N_{\mathrm{r}}}).
\end{align}
Therefore, the requirement of $\hat{\by}_1=\hat{\by}_2$ can be decomposed into $2N_{\mathrm{r}}$ independent equations $\mathrm{sgn}(u - v)  =  \mathrm{sgn}(u + v)$, which gives
\begin{equation}
  \mathrm{Pr}\left(\hat{\by}_1=\hat{\by}_2\right)=\left(\frac{2}{\pi}\arctan\sqrt{\frac{(K-1)P+\sigma^2}{P}}\right)^{2N_{\mathrm{r}}}.
\end{equation}

\bibliographystyle{IEEEtran}
\bibliography{refs_all}

\end{document}